\documentclass[]{spie}  

\usepackage[]{graphicx}

\title{The EBEX Experiment}

\author{P.~{Oxley}\supit{a}, P.~{Ade}\supit{b}, C.~{Baccigalupi}\supit{c}, P.~{deBernardis}\supit{d}, H--M.~{Cho}\supit{e}, M.~J.~{Devlin}\supit{f},
S.~{Hanany}\supit{a}, B.~R.~{Johnson}\supit{a}, T.~{Jones}\supit{a}, A.~T.~{Lee}\supit{e,g}, T.~{Matsumura}\supit{a},  A.~D.~{Miller}\supit{h},
M.~{Milligan}\supit{a}, T.~{Renbarger}\supit{a}, H.~G.~{Spieler}\supit{g}, R.~{Stompor}\supit{i}, G.~S.~{Tucker}\supit{j}, M.~{Zaldarriaga}\supit{k}
\skiplinehalf
\supit{a} Dept. of Physics and Astronomy, University of Minnesota, Minneapolis, MN 55455, U.S.A.; \\
\supit{b} Dept. of Physics and Astronomy, University of Wales, Cardiff, CF24 3YB, Wales; \\
\supit{c} International School for Advanced Studies, 34014 Trieste, Italy; \\
\supit{d} Dipartimento di Fisica, Universita di Roma La Sapienza, 00185 Rome, Italy; \\
\supit{e} Dept. of Physics, University of California, Berkeley, CA 94720, U.S.A.; \\
\supit{f} Dept. of Physics, University of Pennsylvania, Philadelphia, PA 19104, U.S.A.; \\
\supit{g} Physics Division, Lawrence Berkeley Nat. Lab., Berkeley, CA 94720, U.S.A.; \\
\supit{h} Dept. of Physics, Columbia University, New York, NY 10027, U.S.A.; \\
\supit{i} Computational Research Division, Lawrence Berkeley Nat. Lab., Berkeley, CA 94720, U.S.A.; \\
\supit{j} Dept. of Physics, Brown University, Providence, RI 02912, U.S.A; \\
\supit{k} Astronomy Department, Harvard University, Cambridge, MA 02138, U.S.A; \\
}
\authorinfo{Send correspondence to P.~Oxley: E-mail: oxley@physics.umn.edu, Telephone: +1 612 624 0673\\ }

\def\maxipol{MAXIPOL}
\def\hwp{HWP}

\def\igb{IGB}
\def\apex{APEX}
\def\ebex{EBEX}
\def\squid{SQUID}

\def\micron{$\mu$m}
\def\microK{$\mu{\mbox{K}}$}

\def\ruo2{RuO$_{2}$}

\def\degr{$^{\circ}$}

\def\sq{$Q$}
\def\su{$U$}

\def\mathrelfun#1#2{\lower3.6pt\vbox{\baselineskip0pt\lineskip.9pt
  \ialign{$\mathsurround=0pt#1\hfil##\hfil$\crcr#2\crcr\sim\crcr}}}
\def\simlt{\mathrel{\mathpalette\mathrelfun <}}

\long\def\comment#1{}

\newcommand{\wisk}[1]{{\ifmmode{#1}\else{$#1$}\fi}}
  
\begin{document} 
  \pagestyle{plain}
  \setlength{\baselineskip}{0.99\baselineskip}
  \setlength{\parindent}{0pt}
  \maketitle 

\begin{abstract}
EBEX is a balloon-borne polarimeter designed to measure the 
intensity and polarization of the cosmic microwave background 
radiation.  The measurements 
would probe the inflationary epoch that took place shortly 
after the big bang and would significantly improve
constraints on the values of several cosmological parameters. 
EBEX is unique in its broad frequency coverage and in its
ability to provide critical information about the
level of polarized Galactic foregrounds which will be necessary
for all future CMB polarization experiments.

EBEX consists of a 1.5~m Dragone-type telescope that 
provides a resolution of less than 8~arcminutes over four focal 
planes each of 4\degr\ diffraction limited field of view at frequencies up 
to 450~GHz. The experiment is designed to accommodate 330 
transition edge bolometric detectors per focal plane, for 
a total of up to 1320 detectors. EBEX will operate with 
frequency bands centered at 150, 250, 350, and 450~GHz. 
Polarimetry is achieved with a rotating
achromatic half-wave plate.  EBEX is currently in the design and 
construction phase, and first light is scheduled for 2008.
\end{abstract}


\section{Introduction}
EBEX (E and B EXperiment) is a long duration balloon-borne experiment 
equipped with arrays of hundreds of bolometric transition edge sensors
(TES) at the focal plane of a 1.5~m aperture telescope.  It will
measure the temperature and polarization of the cosmic microwave 
background (CMB) radiation.  EBEX has
four primary scientific goals, (1)~to detect the B-mode inflationary
gravitational-wave background signal, or set an upper bound that is
a factor of $\sim$15 more restrictive than current bounds, (2)~to provide
critical information about the polarization of Galactic foregrounds,
particularly of dust emission, at the micro-Kelvin level, (3)~to
measure the yet undetected signature of lensing of the polarization of
the CMB, and (4)~to improve the determination of several cosmological
parameters by up to a factor of six.  All of these goals are of
fundamental importance for physics and astrophysics. 
\ebex\ will provide an important milestone in the implementation and
testing of detector, detector readout, optics, and polarimetry techniques that
are being considered for a future NASA CMB polarization satellite, and is
unique in its foreground determination capability compared to all
current and proposed CMB experiments.
In Sec.~\ref{sec:science} we review the 
cosmology that will be probed with the EBEX measurements and in 
Sec.~\ref{sec:ExpDetails} we describe the EBEX instrument.

\section{Science}
\label{sec:science}

Since its discovery in 1965 the cosmic microwave background radiation has
been one of the pillars of the Big Bang model~\cite{penzias65}.
Measurements of its spectrum firmly established the hot big bang model
as the basis of our understanding of cosmology~\cite{mather94}.
Measurements of the anisotropy of the CMB over the last fifteen years have 
enabled us to determine cosmological parameters such as the age, density and composition of the 
universe~\cite{spergel03,tegmark03} to unprecedented accuracy. Recent measurements 
of the polarization of the CMB provide strong support for our model 
of the CMB~\cite{kovac02,leitch02b}.


As a result of these and other astrophysical observations we have a
very successful model for how structures have formed in our
universe. Gravitational instability makes fluctuations grow with
time, but the fluctuations require initial seeds.  The CMB temperature
power spectrum and the angular dependence of the cross-correlation
between temperature and polarization~\cite{kogut03a} imply that these
seeds had to be created very early in the evolution of the
universe~\cite{hu97c,spergel_z97,peiris03}.  


The leading scenario for the creation of the primordial seeds is the
paradigm of inflation~\cite{guth81,linde82,albrecht82,sato81}.  The
fluctuations we see today are the direct result of quantum
fluctuations of a very light scalar field, the inflaton, during a
period of accelerated expansion or ``inflation'', a small
fraction of a second after the big 
bang~\cite{mukhanov81,hawking82,guth82,starobinsky82,starobinsky83a,bardeen83,mukhanov92}.  
However, many of the details of the
inflationary scenario are uncertain and perhaps  more importantly, 
the paradigm currently lacks any strong confirmation.

\begin{figure}
   \begin{center}
   \begin{tabular}{c}
   \hspace{-0.3in}
   \includegraphics[height=7cm]{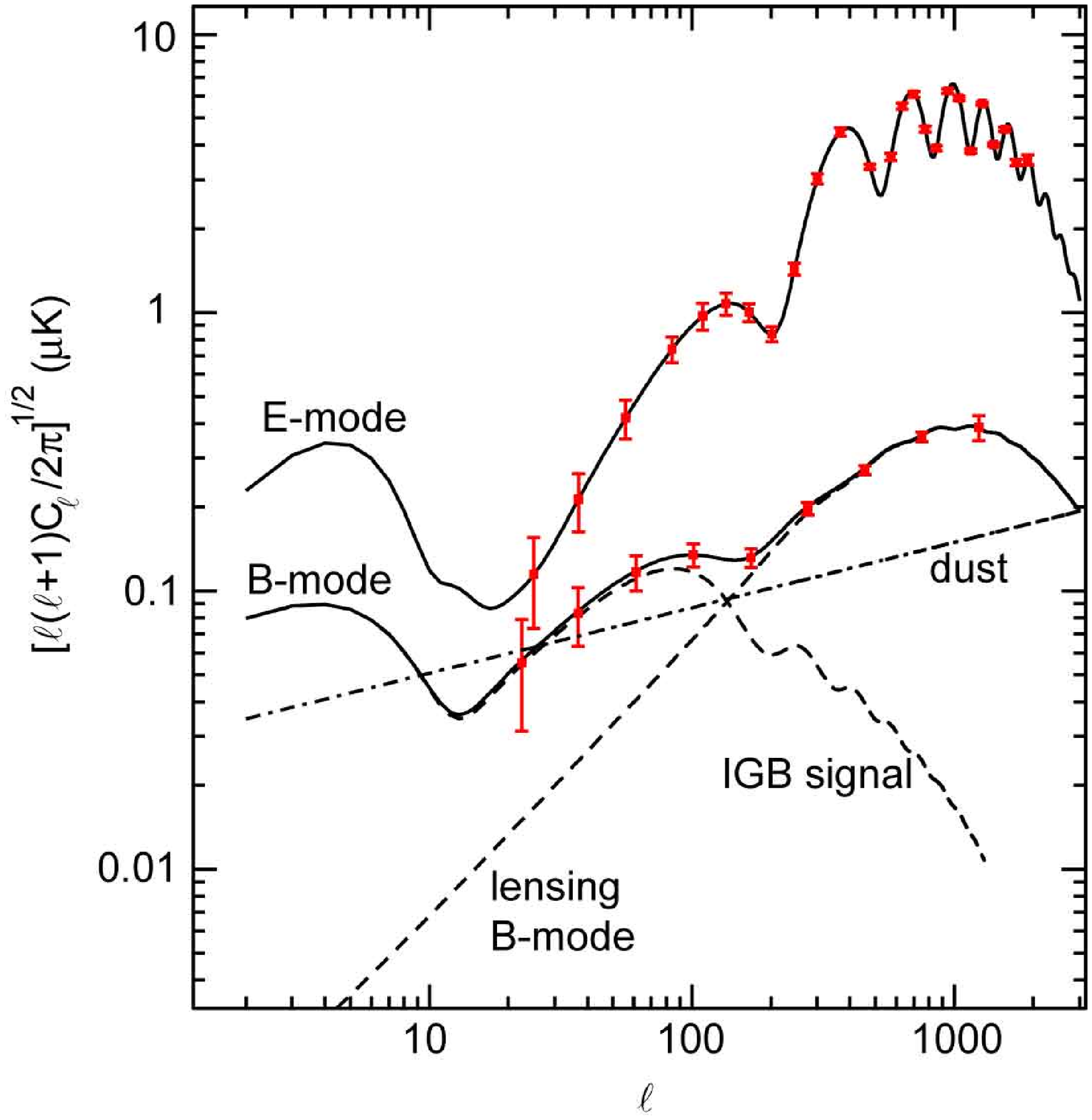}
   \hspace{0.25in}
   \includegraphics[height=6cm]{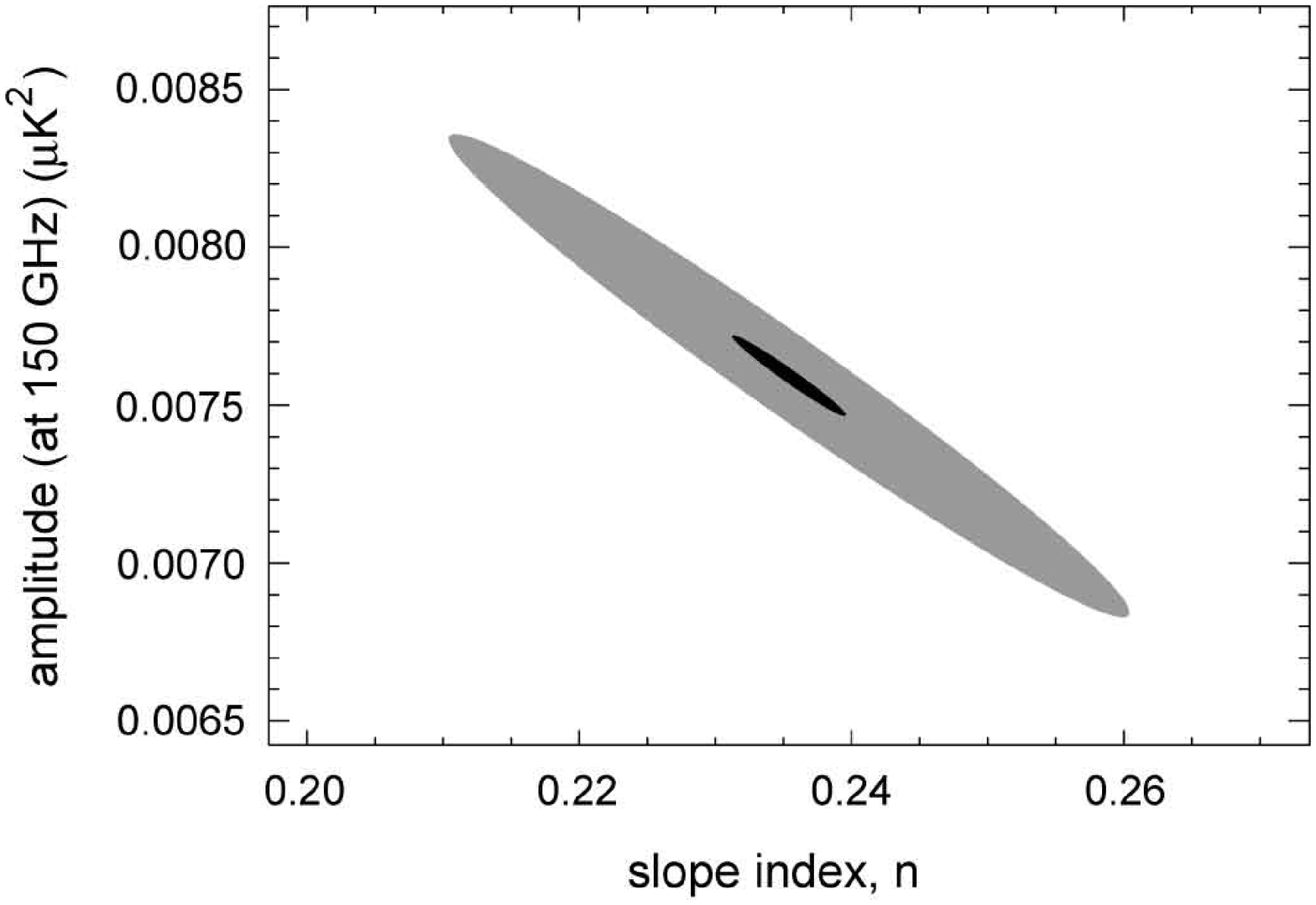}
   \end{tabular}
   \end{center}
   \vspace{-0.1in}
   \caption[science] 
{Left panel:  The angular power spectrum of the CMB E and B-mode signal 
at 150~GHz (solid lines).  The B-mode signal is the sum of contributions from 
lensing and the IGB (dashed lines) in a $\Lambda$CDM cosmological model with
the amplitude of the \igb\ signal a factor of five lower than current upper limits.  
The x-axis is spherical multipole moment $\ell$.
The data points represent the expected performance of EBEX after a 14 day 
flight in Antarctica~\cite{ball}.
Also shown is the estimated dust foreground angular power spectrum 
at 150~GHz for the region of the 
sky to be scanned by EBEX. The dust foreground shown is the
same in both E and B-type polarization patterns, and its amplitude and
slope will be determined by EBEX measurements at 250, 350, and 450~GHz.  
Right panel: 
Two sigma uncertainty ellipses in EBEX determination of the 
amplitude and angular power spectrum slope 
of the 150~GHz B-mode power spectrum from dust.
With 55, 30, and 25 detectors at 250, 350, and 450~GHz, respectively,
and for a 14 day flight, the amplitude and slope will be determined to 
6\% (lighter colored ellipse).  With 330, 180, and 150 detectors at
these frequencies the two sigma uncertainty is 1\% (darker colored ellipse). 
Nominal values of 0.00759~$\mu$K$^2$ and 0.235 for the amplitude
and slope are used~\cite{bacci_private}.}
\label{fig:science}
\end{figure}

A promising way to confirm the inflationary scenario is 
via its prediction of a stochastic background of gravity 
waves~\cite{rubakov82,starobinsky82,starobinsky83a,grishchuk75,abbott84a},
which we call the inflationary gravitational-wave background (IGB).
The best known way to search for the IGB is through its signature on
the CMB polarization~\cite{kamionkowski97b,seljak97}.  
Thomson scattering of an anisotropic radiation field leads
to linear polarization.  The temperature anisotropy produced by
density perturbations and by gravity waves both lead to linear polarization
through this mechanism. However, the pattern of the induced
polarization on the sky should be very different in each case. Density
perturbations produce a curl-free or E-mode pattern of polarization
vectors.  Gravity waves produce a curl or B-mode pattern of
polarization vectors that density perturbations cannot 
produce~\cite{kamionkowski97a,zaldarriaga97}.  
Thus if polarization can be mapped and decomposed
into E and B components (Fig.~\ref{fig:science}, left panel), 
the B component will be an unambiguous
signature of the IGB and therefore for inflation.
Detection of this IGB signal is a primary goal of the EBEX experiment.

The \igb\ signal is extremely faint -- with a similar or smaller
amplitude than expected foreground signals.
It is anticipated that the most significant foreground signal
will be from polarized Galactic 
dust\cite{baccigalupi03} (Fig.~\ref{fig:science}, left panel).
However, there is currently very little information about
the level of polarized dust and its orientation as a function of
position on the sky, or about its E and B power spectra. No
information exists for any region of the sky at the accuracies
required for a B-mode signal detection.  A primary goal of EBEX
will be to characterize the polarized dust emission and determine 
its angular power spectra in both E and B-type polarizations.  

The dust signal increases
with increasing frequency, while the CMB signal is at a maximum
at $\approx$150~GHz.  The measurements at 250, 350, and 450~GHz will be 
primarily sensitive to dust emission, and by extrapolating these measurements
to 150~GHz the dust foreground will be subtracted from the
primary CMB band at 150~GHz.  The predicted uncertainty in the determination
of the dust signal at 150~GHz is shown in the right panel of Fig.~\ref{fig:science}.
The dust measurements will also provide the community with
critical information about the polarized dust fraction and orientation
as a function of frequency. Given the expected magnitude of the
polarized dust foreground, this information is essential for {\it all}
future CMB polarization experiments.

We have chosen a balloon-borne platform for EBEX because this allows
measurements over a broad frequency range and reduces
atmospheric effects by three orders of magnitude~\cite{hanany03b}.
At frequencies higher than
$\sim$150~GHz atmospheric emission becomes significant and broad-band
observations from the ground are difficult because of a combination of
lower detector sensitivity (a result of increased incident power) and
higher noise (due to fluctuations in sky emission). Broad-band,
ground-based observations are essentially impossible from anywhere on
Earth at the higher frequency bands.  EBEX is unique among CMB
polarization experiments in that it will have four frequency bands
between 150 and 450~GHz. This is the broadest frequency coverage
of all current and proposed bolometric CMB polarimeters and gives
EBEX an unprecedented capability to measure the polarization of the dust
emission (Fig.~\ref{fig:science}, right panel).

The atmosphere is a source of polarized intensity because of Zeeman-splitting
of rotational levels of oxygen.  The levels are split
by the Earth's magnetic field and the amount and orientation of the
polarized intensity depends on the strength and direction of the
field~\cite{hanany03b,keating98}.  The atmospheric linear polarization 
is predicted to occur at
only the $\sim$$10^{-9}$~K level, but circular polarization is predicted
at a level of $\sim$100~$\mu$K~\cite{hanany03b}.  A conversion of
circular to linear polarization in any ground-based instrument with a
level as low as 0.1\% would give rise to an instrumental polarization
signal that is approximately as large as the IGB signal of 0.1~$\mu$K
shown in Fig~\ref{fig:science}, left panel.  At 90~GHz the atmospheric 
signal is $\sim$5 times the \igb\ signal.  At balloon altitudes 
the atmospheric signal is a factor of 1000 smaller and therefore should 
be negligible for EBEX.

Gravitational lensing of CMB
photons along their path gives rise to an additional effective
foreground.  The lensing is the result of the deflection of the path of
CMB photons by intermediate mass and energy structures, and it creates a
B-mode polarization pattern even in the absence of gravity
waves~\cite{zaldarriaga98}.
Based on current measurements of cosmological parameters the
magnitude of the lensing B signal is predicted with $\sim$20\%
uncertainty~\cite{spergel03,tegmark03}, but because of its low
amplitude it has not yet been detected.  If the amplitude of the
lensing signal agrees with predictions, EBEX should have the combination
of sensitivity and angular resolution to detect and constrain its
amplitude to within 4\%.  If EBEX achieves its predicted sensitivity a
 ``no-detection'' would be of substantial
importance because it would require major revisions of our
understanding of the evolution of the universe.

EBEX should make a high
signal-to-noise ratio determination of the CMB temperature and 
the E-mode polarization spectrum, and the spectrum of the 
correlation between the two.  The E-mode measurement will allow a factor of 
between two and three improvement in current estimates of the density of dark energy, 
the density of dark matter, the density of baryonic matter, and 
the amplitude of the primordial density fluctuations spectrum. 
The spatial distribution of density
perturbations in the early universe is parameterized by their 
spectral index and EBEX should improve current constraints
on the running of this spectral index by a factor of six.

\section{Experimental Details}
\label{sec:ExpDetails}
	\subsection{Summary of Approach} 
EBEX will employ a 1.5~m telescope to focus radiation on up to four 
separate focal planes, each containing up to 330 TES spider web bolometers
cooled to 300~mK. 
The experiment will operate at four frequency bands, centered on 150,
250, 350, and 450~GHz.  
The unpolarized ($T$) and linearly polarized ($Q$ and $U$) Stokes
parameters of the CMB radiation will be measured by the well-tested
technique of rotating an achromatic half-wave plate at the aperture
stop of the telescope and analyzing the radiation with a wire grid
polarizer. The telescope will scan a patch of sky 350~deg$^2$
with a resolution of 8~arcminutes at 150~GHz (Table~\ref{tab:band_sens}).

\begin{figure}
   \begin{center}
   \begin{tabular}{c}
   \includegraphics[height=9cm]{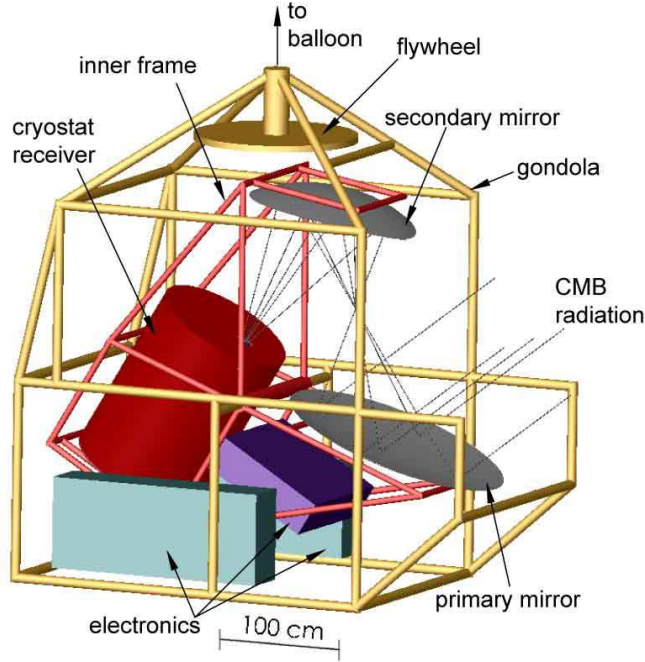}
   \end{tabular}
   \end{center}
   \vspace{-0.1in}
   \caption[telescope] 
{Solid model of the EBEX experiment and balloon gondola showing the telescope 
primary and secondary mirrors and
the cryostat which contains re-imaging optics and the TES detectors.}
\label{fig:ebex_experiment}
\end{figure}

EBEX is carefully optimized to achieve its science goals.  The
balloon-borne platform and the broad frequency coverage will provide
critical information about the dust foreground. The large number of
detectors will give the required combined sensitivity that will allow
detection of the IGB signal.  Figure~\ref{fig:ebex_experiment} shows 
the EBEX payload. 

	\subsubsection{Detectors and Readout}
	\label{sec:detectors_technical}

The detection of the IGB signal requires high sensitivity. However,
detector technologies have matured to the point that the noise of a
bolometric detector is dominated by noise in the background optical loading and
therefore the sensitivity of a single detector cannot be improved. To
improve sensitivity it is essential to implement arrays of detectors.

We have chosen to use arrays of transition edge sensors for EBEX for
several compelling reasons.  They are produced by thin film deposition
and optical lithography, which is ideal for producing large detector
arrays.  Due to a strong negative electrothermal feedback effect, they
are extremely linear and the responsivity is determined only by the bias
voltage, insensitive to fluctuations in the incident optical power and
bath temperature.  The electrothermal feedback also strongly reduces
Johnson and 1/f noise in the TES. TES bolometers are low-impedance
devices, which results in a lower vibration sensitivity than that of
high-impedance semiconductor bolometer technologies.  SQUID amplifiers
will be used to read out the TES bolometers.  These amplifiers are
inherently low noise devices and there is the possibility to use
multiplexed SQUID amplifiers, which will simplify our cryogenic design
and facilitate the use of the full 1320 detector capability of EBEX.

\begin{figure}
   \begin{center}
   \begin{tabular}{c}
   \includegraphics[height=5.5cm]{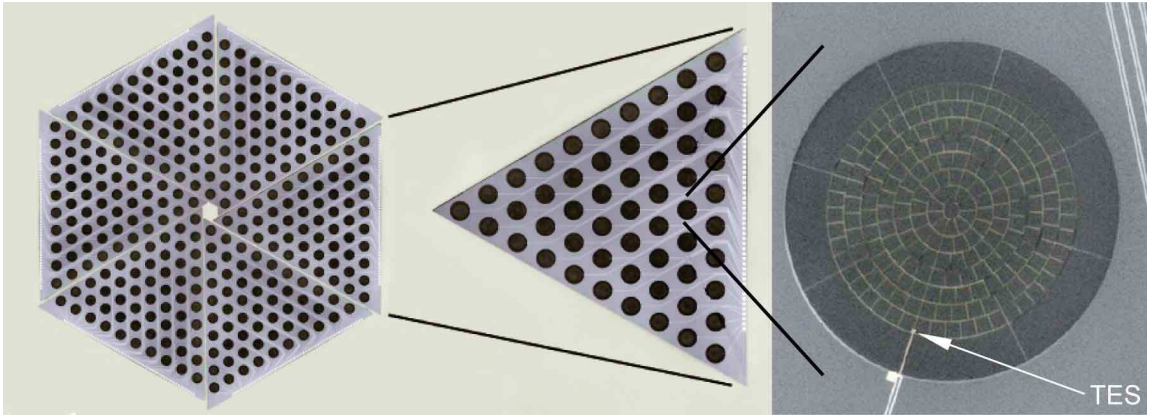}
   \end{tabular}
   \end{center}
   \vspace{-0.1in}
   \caption[TES array] 
{Left panel:
Montage image of a photograph of a single real TES ``wedge'' of spider
bolometers replicated electronically to show how a final array
will look.  Each wedge is designed to be identical and has 55
bolometers.  The entire array will have 330 bolometers and be 17~cm
in diameter.
Center panel:
Photograph of wedge of 55 spider web TES
bolometers.  Wedge is 8~cm on a side.  Suspended spider web absorbers
are fabricated from 1~$\rm{\mu m}$ thick silicon nitride.  The
membrane is released from the front side using a gaseous xenon
diflouride etch.  Bolometers are 3.5~mm diameter with 0.5~mm long legs.
Wiring layer is superconducting aluminum. This array was fabricated in
the U.C. Berkeley microfabrication facility. 
Right panel:
Close up of 55 element bolometer wedge.  Sensors are an Al/Ti
proximity effect sandwich TES, located at the edge of the 
membrane and electrically connected with superconducting leads.}
\label{fig:full_array}
\end{figure}

In our TES spider web bolometer CMB radiation is absorbed by a 
metalized silicon nitride mesh
in a pattern similar to that of a spider's web 
(Fig.~\ref{fig:full_array}, right panel).  
EBEX will use arrays of 55 TES bolometers fabricated by our collaborators 
at UC Berkeley using standard microlithographic techniques. 
A 330 element TES array is built by tiling six 55 element arrays
(Fig.~\ref{fig:full_array}, left panel) and  will fill the EBEX focal plane.
The arrays are identical in many of their
construction parameters to the arrays that will be used
for the ground-based experiment \apex~\cite{schwan03} and for 
the South-Pole Telescope (SPT)~\cite{spt}.  However, compared to the \apex\ TES the
thermal conductance of the EBEX TES will be reduced to
10~pWK$^{-1}$ to match the lower optical loading at
balloon altitudes.

\begin{figure}
   \begin{center}
   \begin{tabular}{c}
   \includegraphics[height=7cm]{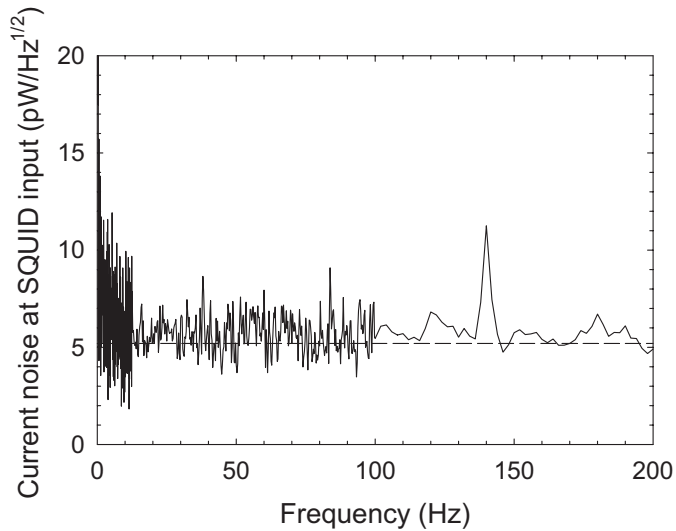}
   \end{tabular}
   \end{center}
   \vspace{-0.1in}
   \caption[TES noise] 
{Noise performance of an individual spider web TES bolometer
with no incident optical power.  
The noise is at the fundamental thermal noise limit as
indicated by the horizontal dashed line, and the time constant
is 0.6~ms. The feature at 140~Hz is a mechanical resonance of 
the table upon which the bolometer is mounted.}
\label{fig:noise}
\end{figure}

The responsivity, noise properties and time constants of individual
TES bolometers have been studied
extensively~\cite{lee96,lee98,gildemeister99,gildemeister00,gildemeister00b}.
The measured noise has been shown to reach the fundamental
thermal-fluctuation noise limit (Fig.~\ref{fig:noise}), and the
responsivities and time constants match theory closely.  
Table~\ref{tab:band_sens} gives the expected noise equivalent power
(NEP) of the EBEX detectors at each frequency.

Initially EBEX will use individual SQUIDs to read the current from each
TES bolometer using the readout technique that has been developed for
the APEX and SPT receivers.  This technique uses a single series-array
SQUID coupled to a single room-temperature op-amp.  The TES
bolometer is ac-biased with a direct-digital synthesis generator and
the signal is demodulated with an analog mixer circuit. 
Since SQUIDs are sensitive to stray magnetic fields the TES and SQUID
will be enclosed in a Cryoperm shield.

The above readout system is fully compatible with the
frequency-domain multiplexing readout that is being
developed by our collaborators at UC Berkeley and Lawrence Berkeley National
Lab.  The principle of this
technique has already been demonstrated by monitoring the current
passing through 8 mock TES~\cite{yoon01}.  Two real TES have also been
multiplexed~\cite{cunningham02} and experiments are underway to
further test the technique with greater numbers of TES bolometers.
With multiplexed readouts we will use one \squid\ readout circuit to
read 20 detectors, thereby reducing the number of SQUID wires reaching
cryogenic temperatures by the same factor.  This substantially
reduces the heat load on the cryostat allowing
use of the full 1320 detector capability of EBEX.  When individual
SQUIDs are used to read out each TES the number of detectors will be
limited to 715, distributed between the four focal planes.  
Multiplexed readouts will allow more detectors to be allocated to 
the 250, 350, and 450~GHz frequency bands, improving the determination 
of the dust foreground compared to the case of non-multiplexed 
readouts (Fig.~\ref{fig:science}, right panel).

The TES bolometers will operate from a bath temperature of $\sim$300~mK
provided by a $^3$He fridge.
The fridge will work off of a standard liquid helium/liquid nitrogen
cryostat with two vapor cooled shields connected to the
boil-off gases of each of the liquids. 
A detailed thermal model of the cryostat has been made, 
including all anticipated major heat loads such as radiation, wires,
heat load through the 15~cm diameter cryostat window, and parasitic
heat loads on the different thermal stages.  The modelling predicts that
105~l of liquid helium and 140~l of liquid nitrogen will be sufficient
for a hold time of 20~days.

	\subsubsection{Telescope and Cold Optics}
	\label{sec:optics_technical}

\begin{figure}
   \begin{center}
   \begin{tabular}{c}
   \hspace{-0.2in}
   \includegraphics[height=7cm]{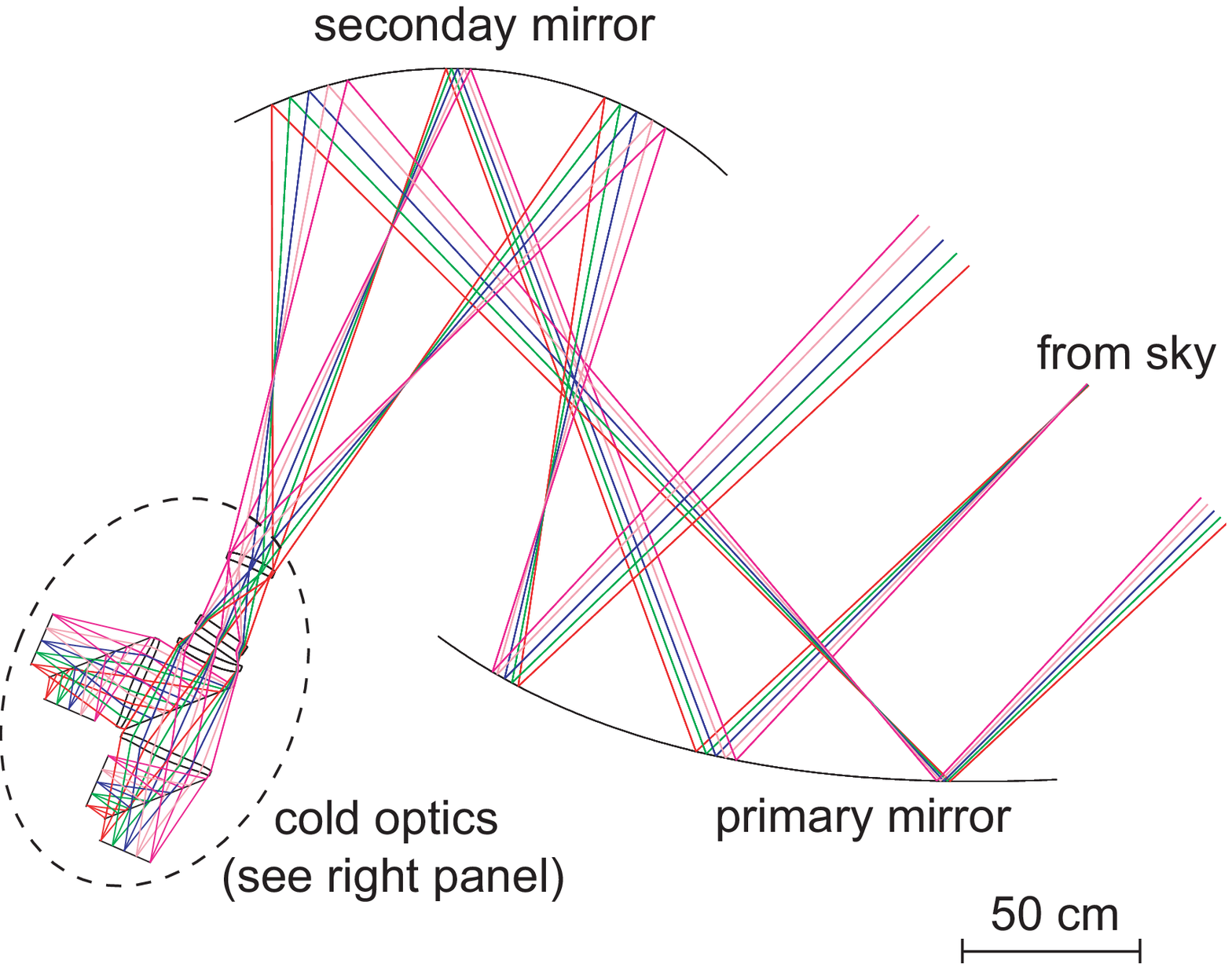}
   \hspace{0.2in}
   \includegraphics[height=7cm]{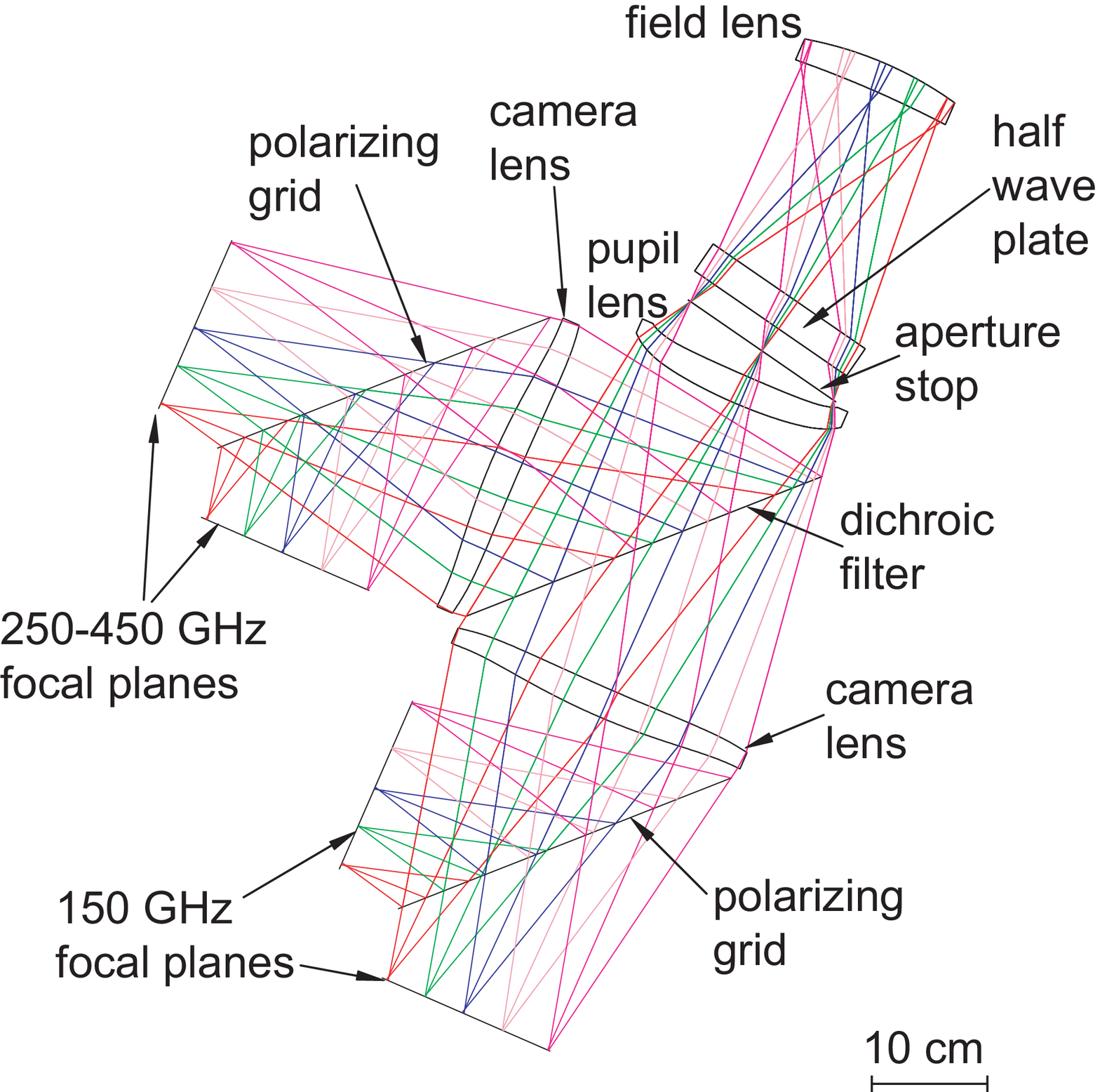}
   \end{tabular}
   \end{center}
   \vspace{-0.1in}
   \caption[telescope] 
{Left panel:  The 1.5~m Dragone telescope provides 8~arcminute beams at 150~GHz.  Light
is coupled from the telescope to the cold optics, seen in close up on the right. 
Right panel:  The cold optics image two perpendicular polarization components from 
each of the four frequencies onto four focal planes.}
\label{fig:telescope}
\end{figure}

The EBEX optical system is based on the existing well-characterized
telescope from the Archeops experiment. It consists of the Archeops 
1.5~m aperture primary mirror and a new 1.2~m elliptical secondary that
together make a Dragone telescope (Fig.~\ref{fig:telescope}, left panel). 
This design provides a 4\degr\ field of view and   
beam resolutions given in Table~\ref{tab:band_sens}.
\begin{table}
\caption{
Expected TES noise and angular resolution for EBEX. 
The TES noise is dominated by photon and thermal noise.
Photon noise is due to fluctuations in the number of
CMB photons arriving at the TES and thermal noise is caused by
fluctuations in the number of phonons in the heat link between
the TES and the low temperature bath used to cool the TES.
The angular resolution is determined by the telescope 1.5~m
aperture primary mirror.}
\begin{center}
\begin{tabular}{ccc} \hline \hline
Frequency & NEP & Angular Resolution \\
(GHz) & $\times 10^{-17}$~WHz$^{-1/2}$ & (FWHM, arcminutes) \\  \hline
150 & 1.12 & 8 \\
250 & 1.28 & 4.8  \\ 
350 & 1.48 & 3.4  \\ 
450 & 1.71 & 2.7  \\ \hline
\end{tabular}
\vspace{-0.05in}
\label{tab:band_sens}
\end{center}
\vspace{-0.15in}
\end{table}

Figure~\ref{fig:telescope}, right panel, shows the optics
of EBEX contained within the cryostat receiver of Fig.~\ref{fig:ebex_experiment}.
Four cold lenses, a dichroic filter and two polarizing grids
couple radiation from the telescope into four independent focal
planes. The lenses, which will be made of either silicon or high
density polyethylene, provide both a cold aperture stop, where the
rotating half-wave plate is mounted, and a flat and telecentric 
focal plane suitable for the TES arrays.  
This configuration provides a 4\degr\ field
of view for each focal plane with Strehl ratios of at least 0.966 at
all frequencies across the entire field of view, and a Strehl ratio
better than 0.991 across the full field of view for 150~GHz.
To enable CMB observations to be made in four distinct frequency bands
the incident light will be filtered, and different frequencies fed
to different parts of the four focal planes.
A reflecting metal-mesh low-pass filter with an
edge at $\sim$510~GHz will be mounted near the aperture stop of the
system. The dichroic filter has an edge at $\sim$210~GHz, and acts as
a low-pass filter in transmission
and a high-pass filter in reflection. Two focal planes with up to a
total of 660 detectors that are fed by light transmitted by this
filter are allocated for the 150~GHz detectors. The other two focal
planes with up to a total of 660 detectors are allocated for the three
higher frequency bands. For each of the frequency bands, bandpass
filters mounted in front of the focal plane will define the exact
band shape.  The capacitive inductive meshes of these filters are
embedded in a transmissive polymer giving them mechanical rigidity
even when cut to non-circular shapes. This flexibility allows us to
use different parts of a given focal plane for different frequency
bands.

Both silicon and high density polyethylene (HDPE) are commercially
available at the sizes required for the lenses. Silicon and HDPE have
indices of refraction of 3.416 and 1.567 at 
mm-wavelengths~\cite{birch84,birch93,parshin95,afsar94}, which
would give rise to coefficients of reflection of 30\% and 4\%, and to
an absorption of 8\% and 23\% respectively at 150~GHz.  
We plan to coat either type of lens with an
anti-reflection coating but the challenge is to develop a broad-band
coating that adheres to the lens without causing
damage when cooled to cryogenic temperatures.   
There is great interest in solving this problem
in the larger CMB community because many experiments, including ACT,
SPT, QUIET, BICEP, and perhaps a NASA CMB polarimetry satellite will
need this technology.

Light will be coupled to the TES array by an array of smooth-walled
conical horns.  The use of a horn
array places the detectors inside a Faraday cage where they are
shielded from electromagnetic interference which may enter the
cryostat window.  The horns will have $1.7\,F\,\lambda$ diameter entrance
apertures at 150~GHz which allows 330 detectors to fill the
4\degr\ field-of-view of the EBEX telescope.  The array will be fed
at $F=1.9$, giving a $6.5$~mm horn aperture at 150~GHz, closely matching
the bolometer spacing in the TES array.

The horn array will be made in two parts.  A top section which is an
array of conical holes, and a lower section an array of cylindrical
holes acting as waveguides.  Sandwiched between the two is a Cryoperm
plate, also with holes in it (Fig.~\ref{fig:ahwp_horns}, left panel).  This plate
is the top plate of a large Cryoperm box which encloses the TES array and
the SQUID detectors, which are located inside an additional smaller
Cryoperm box at 4~K on the cryostat cold plate below the TES array.
Each box attenuates external magnetic fields by more than two orders
of magnitude which is necessary to avoid spurious signals in the SQUIDs
and TES due to magnetic pickup from the half-wave plate rotation mechanism 
(Sec.~\ref{sec:polarimetry_technical}).  We calculate that with this magnetic
shielding the spurious signal induced in the \squid\ will be
a factor of 300 smaller than the
level of detector noise, while the signal in the TES
is an additional factor of 2 smaller.  
The horn array and large Cryoperm box are connected to the
4~K cold plate of the cryostat and separated by a small gap from the
TES array at 300~mK.

	\subsubsection{Polarimetry}
	\label{sec:polarimetry_technical}

EBEX will use a combination of a rotating half-wave plate (\hwp) and a
fixed polarizing grid to modulate the polarization signal because this
technique provides strong discrimination against systematic errors. To
date all successful non-interferometric measurements of IR and mm-wave
polarization have used a \hwp\ to modulate
the polarization signal. Examples include the Minnesota Infrared
Polarimeter~\cite{jones88}, Stokes~\cite{platt91},
Millipol~\cite{leach91}, and SCUBA~\cite{murray97}.  The majority 
of bolometric CMB polarimeters will use a 
\hwp~\cite{church03b,johnson03b,polarbear}. 

\begin{figure}
   \begin{center}
   \begin{tabular}{c}
   \includegraphics[height=5cm]{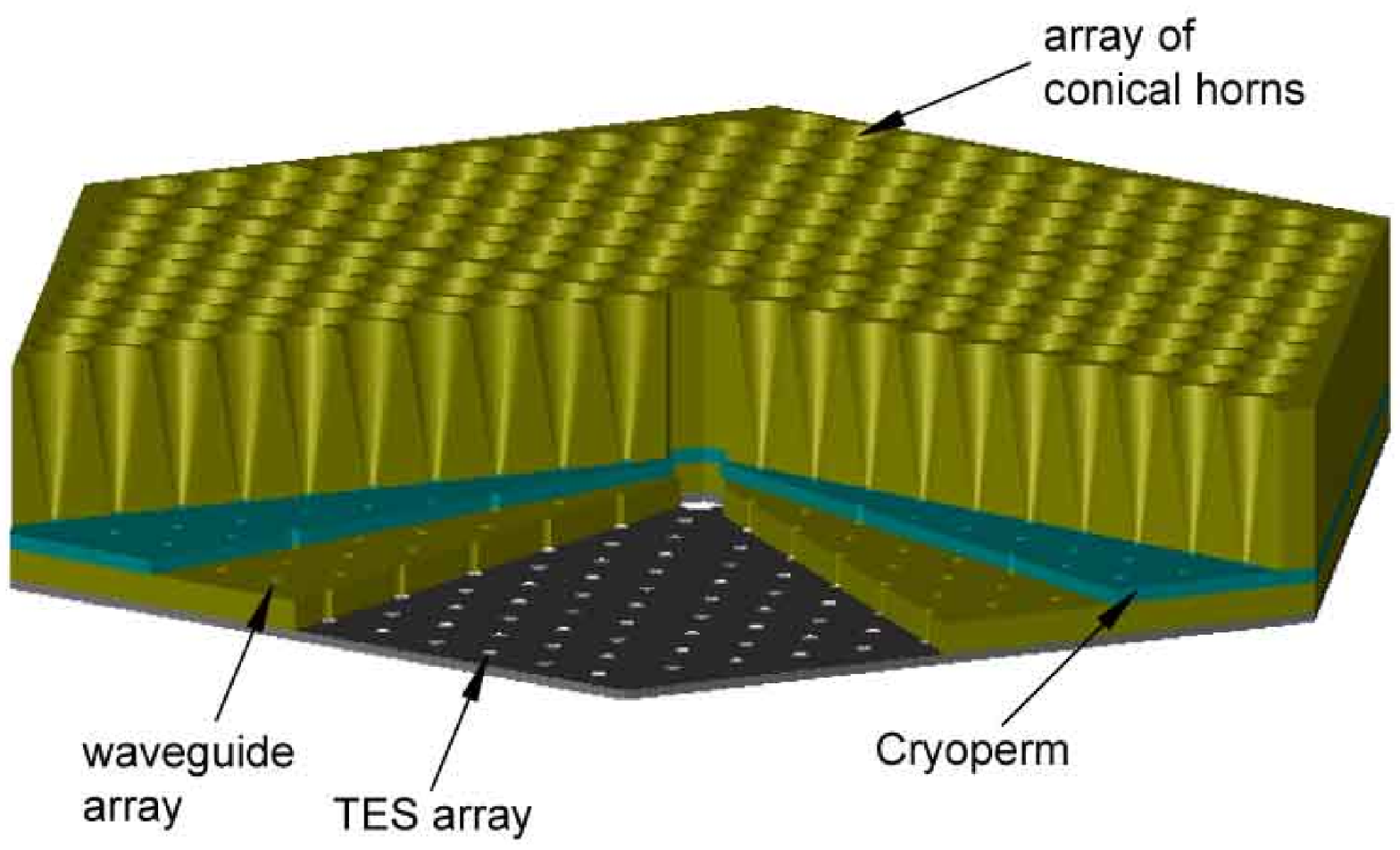}
   \hspace{0.3in}
   \includegraphics[height=5cm]{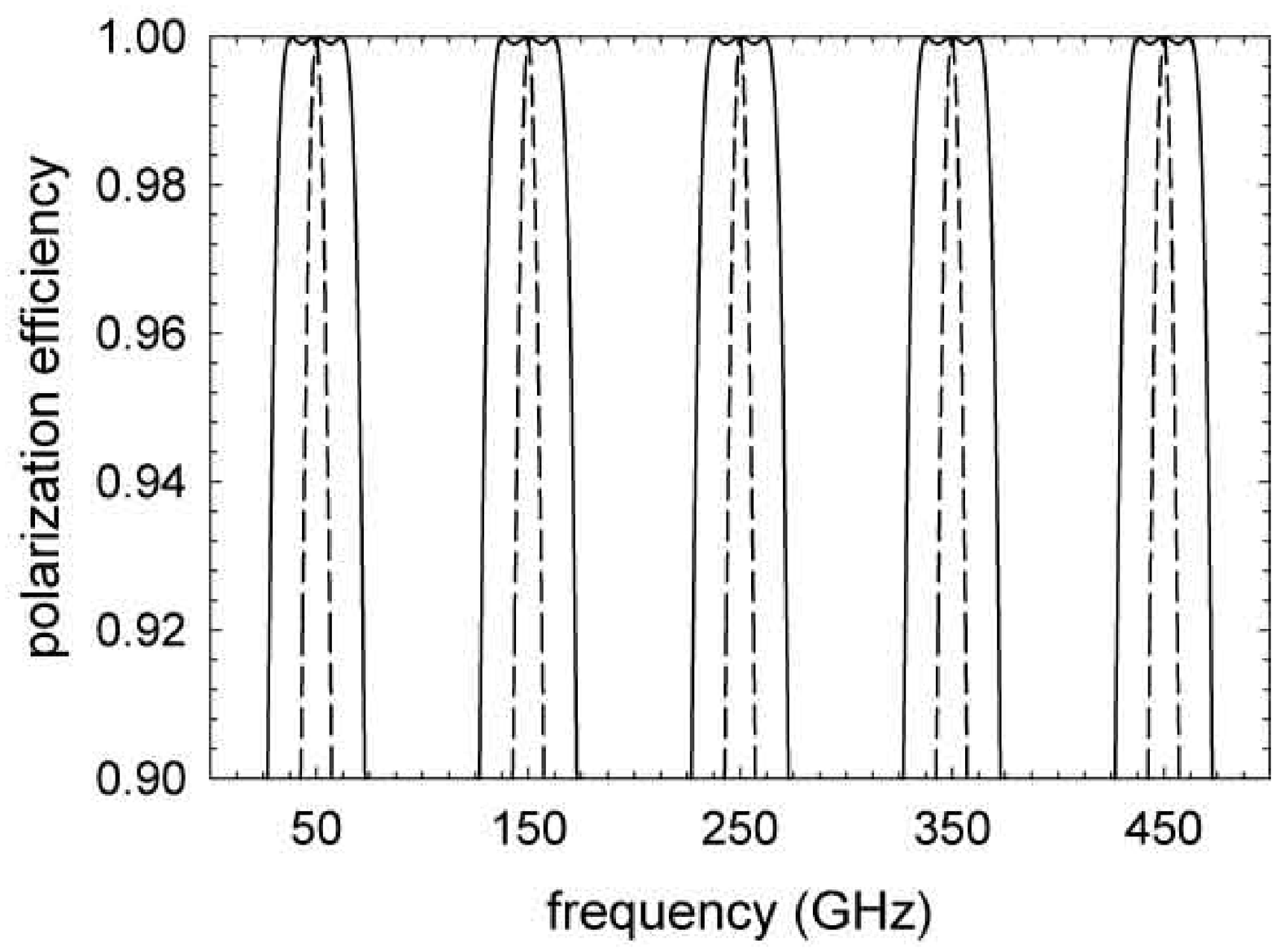}
   \end{tabular}
   \end{center}
   \vspace{-0.1in}
   \caption[awhp] 
{Left panel:  Coupling light to the TES array
with an array of conical horns and waveguides separated by
a high permeability Cryoperm magnetic shield.  The horns, shield,
and waveguides are held at 4~K and the TES array at 300~mK. 
Right panel: Predicted efficiency of a stack of one (dashed line)
and three (solid line) \hwp\ as a function of
frequency. For the 2.84~cm thick A\hwp, each 0.946~cm plate is 
optimized as a single HWP for a
frequency of 50~GHz. The efficiency function has the same
shape and therefore the same bandwidth near the odd-harmonic
peaks of 50~GHz.}
\label{fig:ahwp_horns}
\end{figure}

In EBEX the \hwp\ will be rotated at a frequency of $f_{0}=20$~Hz
which will modulate the polarization at $4 f_0 = 80$~Hz.  The
combination of scanning the sky with the telescope at 1.8\degr s$^{-1}$ and
the rotation of the \hwp\ puts the signal in 15~Hz wide sidebands around
$4f_{0}$. This frequency is well below the $\sim$160~Hz 3~dB
bandwidth of the TES with a 1~ms time
constant.  We will maintain a polarization efficiency that is larger
than 95\% over a 40~GHz bandwidth for each of the four bands by
employing an achromatic \hwp\ (AHWP). The AHWP is made of a stack of
three \hwp's the middle one of which is rotated by 57.5\degr\ with respect 
to the aligned outer plates. The thickness of each \hwp\ is 
optimized for a frequency
of 50~GHz.  Figure.~\ref{fig:ahwp_horns}, right panel, shows the predicted polarization
efficiency for such a stack.  Achromatic \hwp's have been used
in the past at the IR and optical wavelengths~\cite{jones88,koester59,pancharatnam55}.

The use of a HWP as a polarization modulation provides the following
important advantages in discrimination against systematic errors:
\newline 
$\bullet$ Each detector makes independent measurements of $T$,
\sq, and \su\ Stokes parameters for each pixel on the sky.  No
detector differencing is required. All schemes that rely on detector
differencing are prone to errors or increased noise arising from, (a)
uncertainty in the difference in gain between the detectors, (b)
differences in beam pattern (if the detectors do not share the same
light train), (c) difference in absolute calibration, and (d)
difference in noise level between the detectors.  \newline
$\bullet$ Signals reside in relatively narrow and well defined 
sidebands around 4$f_{0}$, where
$f_{0}$ is the HWP modulation frequency. Spurious signals at all other
frequency bands, including spurious effects that might arise from
modulation at $f_{0}$, are rejected. \newline
$\bullet$ Sky signals are constrained to a band of frequencies
that is much higher than the expected $\simlt$1~Hz knee in the $1/f$
noise spectrum of the detectors. \newline
$\bullet$ Sources of instrumental polarization that are on the
detector side of the \hwp\ give rise to systematic effects that
do not modulate with the \hwp\ rotation and therefore do not affect the 
signals from the sky. \newline
$\bullet$ Sources of instrumental polarization that are on the
sky side of the \hwp\ are very stable. Therefore they give rise to a
stable signal at 80~Hz. A notch filter with a width of 
$\sim$0.1~Hz removes such signals without affecting the signals
from the sky. 

\begin{figure}
   \begin{center}
   \begin{tabular}{c}
   \includegraphics[height=7cm]{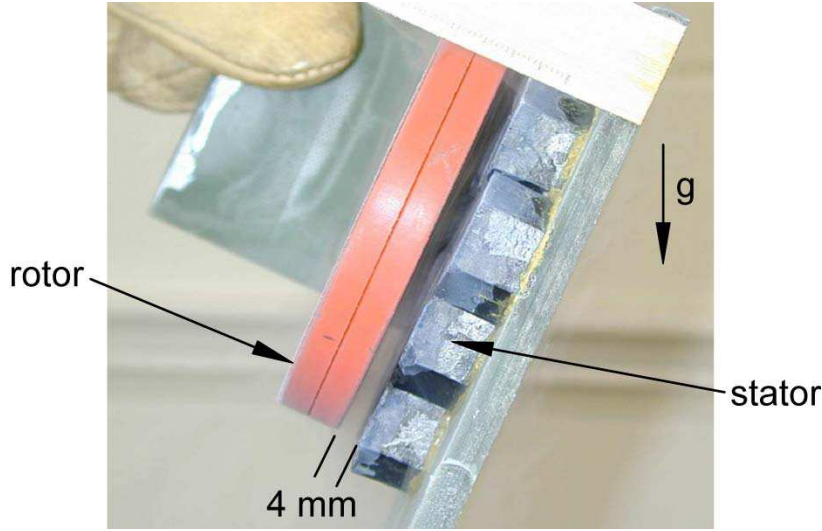}
   \end{tabular}
   \end{center}
   \vspace{-0.1in}
   \caption[SMB in action] 
{The superconducting magnetic bearing consists of a ring shaped
magnet (rotor) and tiles of YBCO (stator) which are
superconducting at temperatures below $\sim$90~K.
The \hwp\ (not shown) is mounted inside the ring magnet.
At temperatures above $T_{c}$ the rotor is held by
mechanical means above the stator. At temperatures below
$T_{c}$ the mechanical constraints are removed and the rotor
maintains its position with respect to the stator {\it
irrespective of the orientation of the acceleration vector due
to gravity}. In this picture the stator was cooled to liquid
nitrogen temperature with the rotor mechanically spaced 
$\sim$4~mm above the tiles. Once cooled the spacing
constraints were removed, the configuration lifted from
the liquid nitrogen and the rotor was set spinning. 
The bearing was then abruptly
tilted by $\sim$70\degr\ with respect to the gravity vector.
No change in the relative orientation of the rotor and stator
and no oscillations of the rotor were observed.  Stiff spring
constants of $2.0\times10^{3}$~Nm$^{-1}$ and $4.2\times10^{3}$~Nm$^{-1}$ 
have been measured in the radial and axial directions, and the oscillation
amplitude at the SMB characteristic frequency is less than 10~\micron~\cite{hanany03}.}
\label{fig:smb}
\end{figure}

We are currently investigating a number of options for rotating the \hwp.
One option is to mount the \hwp\ on a high-temperature
superconducting magnetic bearing (SMB) turned with an induction
motor (Fig.~\ref{fig:smb}).  Bolometers are sensitive to microphonic
pickup and the SMB technique completely eliminates
stick-slip friction, which is the primary source of microphonic noise
in mechanical bearings.  SMBs have been operating for many
years~\cite{rivetti87,decher93,hull97,zhang02} and have been shown 
to reduce friction in bearings by more than 4 orders of magnitude.
We plan to measure the noise properties
and systematic effects associated with the SMB rotation mechanism.
If the SMB is shown to be unsuitable then we have already demonstrated 
with the \maxipol\ experiment that through the use of special
materials, such as vespel, teflon, or rulon, it is possible to reduce
microphonic noise associated with
mechanical bearings to the level appropriate for measurements of the
E-mode polarization~\cite{johnson03b}. Since TES should have
substantially less sensitivity to microphonic noise
than thermistor bolometers~\cite{schwan03} used in \maxipol,
mechanical bearings may be suitable for a B-mode polarization 
experiment such as EBEX.  In the unlikely event that neither 
of these techniques is suitable we will simply 
step-and-integrate with the \hwp.

\subsection{Control of Systematic Errors} 
\label{sec:systematics}

The B-mode signal either from inflation or from lensing is expected to
appear at the $\sim$0.1~\microK\ level at a maximum (Fig.~\ref{fig:science}, left panel)
and therefore a command of the sources of systematic errors is critical. 
We have checked that \ebex\ can meet specific 
goals that are necessary for extraction of
the B-mode signal. Hu et al.~\cite{hu03} have set benchmarks for the
level of residual systematic errors in an experiment attempting to
extract a B-mode signal with a given amplitude. The benchmarks
calculated for EBEX for detecting an amplitude that is {\it a factor
of five lower} than EBEX's $2\sigma$ upper limit are given in
Table~\ref{tab:benchmarks}.  EBEX calibration uncertainty in Stokes
$Q$ and $U$ parameters is expected
to be 1\% -- lower than the 6\% benchmark, and we expect a maximum level
of $QU$ mixing of 0.8\% at the edge of the focal plane.
We expect a pointing uncertainty that is
about a factor of $\sim$3 smaller than the benchmark for 8~arcminute
beams.  In-flight measurements of
Venus (Jupiter) will provide the shape parameters of the beam at each
polarization state to an accuracy of 0.2\% (0.1\%) of the FWHM of the
beam.  We have designed
these planet scans to achieve the benchmark required by 
the monopole term of Table~\ref{tab:benchmarks}. 
The dipole and
differential ellipticity benchmarks are less restrictive and will
already be satisfied by the planet scans.
\begin{table}
\caption{
Types of systematic uncertainties and residual uncertainty
benchmarks for EBEX in order to detect a B-mode signal that is
five times smaller than EBEX $2\sigma$ upper limit (following Hu
et al.~\cite{hu03}).  These benchmarks represent residual
uncertainties after all known corrections have been applied.
The calibration term is the accuracy with which the TES
bolometer response must be calibrated, and the $QU$ term
sets a limit on the allowable level of mixing between
the $Q$ and $U$ polarization states caused by the telescope and optics. 
The pointing, monopole and dipole terms correspond to
uncertainties in the pointing, in the size of beams in two
orthogonal polarization states, and in the alignment of these
beams, respectively, expressed as a fraction of the beam size.
The quadrupole corresponds to uncertainty in the beam
differential ellipticity. }
\begin{center}
\begin{tabular}{|c|c|c|c|c|c|c|} \hline \hline
Type & Calibration & $QU$ & Pointing & Monopole & Dipole & Quadrupole \\ 
\hline
Unit & (\%)   & (\%) & \multicolumn{3}{c|}{fraction of beam} & diff. ell. \\
\hline
Level & 6.4        & 1.6 & 0.11   & 0.002      & 0.007  & 0.017 \\ 
\hline \hline
\end{tabular}
\vspace{-0.05in}
\label{tab:benchmarks} 
\end{center}
\vspace{-0.15in}
\end{table}

\section{Conclusions}
EBEX will measure the CMB temperature and polarization anisotropy on angular 
scales from 8~arcminutes
to 15\degr, thus finding, or setting a substantially improved limit, on the
IGB signal of inflation.  To achieve this EBEX will accurately characterize 
the foreground signal from Galactic dust which will also provide the
CMB community with
critical information about the polarized dust fraction and orientation
as a function of frequency.  EBEX should also detect the
predicted, but as yet undetected, gravitational lensing of CMB radiation and
significantly improve measurements of many cosmological parameters.

\bibliography{proposal}   

\begin{thebibliography}{10}

\bibitem{penzias65}
A.~A. {Penzias} and R.~W. {Wilson}, ``{A Measurement of Excess Antenna
  Temperature at 4080 Mc/s.},'' {\em \apj} {\bf 142}, pp.~419--421, July 1965.

\bibitem{mather94}
J.~C. {Mather}, E.~S. {Cheng}, D.~A. {Cottingham}, R.~E. {Eplee}, D.~J.
  {Fixsen}, T.~{Hewagama}, R.~B. {Isaacman}, K.~A. {Jensen}, S.~S. {Meyer},
  P.~D. {Noerdlinger}, S.~M. {Read}, L.~P. {Rosen}, R.~A. {Shafer}, E.~L.
  {Wright}, C.~L. {Bennett}, N.~W. {Boggess}, M.~G. {Hauser}, T.~{Kelsall},
  S.~H. {Moseley}, R.~F. {Silverberg}, G.~F. {Smoot}, R.~{Weiss}, and D.~T.
  {Wilkinson}, ``{Measurement of the cosmic microwave background spectrum by
  the COBE FIRAS instrument},'' {\em \apj} {\bf 420}, pp.~439--444, Jan. 1994.

\bibitem{spergel03}
D.~N. {Spergel}, L.~{Verde}, H.~V. {Peiris}, E.~{Komatsu}, M.~R. {Nolta}, C.~L.
  {Bennett}, M.~{Halpern}, G.~{Hinshaw}, N.~{Jarosik}, A.~{Kogut}, M.~{Limon},
  S.~S. {Meyer}, L.~{Page}, G.~S. {Tucker}, J.~L. {Weiland}, E.~{Wollack}, and
  E.~L. {Wright}, ``{First-Year Wilkinson Microwave Anisotropy Probe (WMAP)
  Observations: Determination of Cosmological Parameters},'' {\em \apjs} {\bf
  148}, pp.~175--194, Sept. 2003.

\bibitem{tegmark03}
M.~{Tegmark}, M.~{Strauss}, M.~{Blanton}, K.~{Abazajian}, S.~{Dodelson},
  H.~{Sandvik}, X.~{Wang}, D.~{Weinberg}, I.~{Zehavi}, N.~{Bahcall},
  F.~{Hoyle}, D.~{Schlegel}, R.~{Scoccimarro}, M.~{Vogeley}, A.~{Berlind},
  T.~{Budavari}, A.~{Connolly}, D.~{Eisenstein}, D.~{Finkbeiner}, J.~{Frieman},
  J.~{Gunn}, L.~{Hui}, B.~{Jain}, D.~{Johnston}, S.~{Kent}, H.~{Lin},
  R.~{Nakajima}, R.~{Nichol}, J.~{Ostriker}, A.~{Pope}, R.~{Scranton},
  U.~{Seljak}, R.~{Sheth}, A.~{Stebbins}, A.~{Szalay}, I.~{Szapudi}, Y.~{Xu},
  and .~{others}, ``{Cosmological parameters from SDSS and WMAP},'' {\em ArXiv
  Astrophysics e-prints} , Oct. 2003.
\newblock astro-ph/0310723.

\bibitem{kovac02}
J.~M. Kovac, E.~M. Leitch, C.~Pryke, J.~E. Carlstrom, N.~W. Halverson, and
  W.~L. Holzapfel, ``Detection of polarization in the cosmic microwave
  background using {DASI},'' {\em Nature} {\bf 420}, p.~772, December 2002.
\newblock astro-ph/0209478.

\bibitem{leitch02b}
E.~M. Leitch, J.~M. Kovac, C.~Pryke, J.~E. Carlstrom, N.~W. Halverson, W.~L.
  Holzapfel, B.~Reddall, and E.~S. Sandberg, ``Measurement of polarization with
  the degree angular scale interferometer,'' {\em Nature} {\bf 420}, p.~763,
  December 2002.
\newblock astro-ph/0209476.

\bibitem{kogut03a}
A.~{Kogut}, D.~N. {Spergel}, C.~{Barnes}, C.~L. {Bennett}, M.~{Halpern},
  G.~{Hinshaw}, N.~{Jarosik}, M.~{Limon}, S.~S. {Meyer}, L.~{Page}, G.~S.
  {Tucker}, E.~{Wollack}, and E.~L. {Wright}, ``{First-Year Wilkinson Microwave
  Anisotropy Probe (WMAP) Observations: Temperature-Polarization
  Correlation},'' {\em \apjs} {\bf 148}, pp.~161--173, Sept. 2003.
\newblock astro-ph/0302213.

\bibitem{hu97c}
W.~{Hu} and M.~{White}, ``A new test of inflation,'' {\em \prl} {\bf 77},
  p.~1687, Aug. 1996.

\bibitem{spergel_z97}
D.~N. Spergel and M.~Zaldarriaga, ``Cosmic microwave background polarization as
  a direct test of inflation,'' {\em Phys. Rev. Lett.} {\bf 79}, p.~2180, 1997.

\bibitem{peiris03}
H.~V. {Peiris}, E.~{Komatsu}, L.~{Verde}, D.~N. {Spergel}, C.~L. {Bennett},
  M.~{Halpern}, G.~{Hinshaw}, N.~{Jarosik}, A.~{Kogut}, M.~{Limon}, S.~S.
  {Meyer}, L.~{Page}, G.~S. {Tucker}, W.~E., and W.~E. L., ``{First-Year
  Wilkinson Microwave Anisotropy Probe (WMAP) Observations: Implications For
  Inflation},'' {\em \apjs} {\bf 148}, pp.~213--232, 2003.

\bibitem{guth81}
A.~H. {Guth}, ``{Inflationary universe: A possible solution to the horizon and
  flatness problems},'' {\em \prd} {\bf 23}, pp.~347--356, Jan. 1981.

\bibitem{linde82}
A.~D. Linde, ``A new inflationary universe scenario: A possible solution of the
  horizon, flatness, homogeneity, isotropy and primordial monopole problems,''
  {\em Phys. Lett.} {\bf B108}, pp.~389--393, 1982.

\bibitem{albrecht82}
A.~Albrecht and P.~J. Steinhardt, ``Cosmology for grand unified theories with
  radiatively induced symmetry breaking,'' {\em Phys. Rev. Lett.} {\bf 48},
  pp.~1220--1223, 1982.

\bibitem{sato81}
K.~{Sato}, ``{First-order phase transition of a vacuum and the expansion of the
  Universe},'' {\em \mnras} {\bf 195}, pp.~467--479, May 1981.

\bibitem{mukhanov81}
V.~F. {Mukhanov} and G.~V. {Chibisov}, ``{Quantum fluctuations and a
  nonsingular universe},'' {\em JETP Letters} {\bf 33}, pp.~532--535, May 1981.

\bibitem{hawking82}
S.~W. {Hawking}, ``The development of irregularities in a single bubble
  inflationary universe,'' {\em Phys.\ Lett.\ B} {\bf 115}, pp.~295--297, 1982.

\bibitem{guth82}
A.~H. {Guth} and S.~{Pi}, ``Fluctuations in the new inflationary universe,''
  {\em \prl} {\bf 49}, pp.~1110--1113, Oct. 1982.

\bibitem{starobinsky82}
A.~A. Starobinsky, ``Dynamics of phase transition in the new inflationary
  universe scenario and generation of perturbations,'' {\em Phys. Lett.} {\bf
  B117}, pp.~175--178, 1982.

\bibitem{starobinsky83a}
A.~A. {Starobinskii}, ``{The perturbation spectrum evolving from a nonsingular
  initially De-Sitter cosmology and the microwave background anisotropy},''
  {\em Soviet Astronomy Letters} {\bf 9}, pp.~302--+, June 1983.

\bibitem{bardeen83}
J.~M. {Bardeen}, P.~J. {Steinhardt}, and M.~S. {Turner}, ``Spontaneous creation
  of almost scale-free density perturbations in an inflationary universe,''
  {\em \prd} {\bf 28}, p.~679, Aug. 1983.

\bibitem{mukhanov92}
V.~F. {Mukhanov}, F.~A. {Feldman}, and R.~H. {Brandenberger}, ``{Theory of
  cosmological perturbations},'' {\em \phr} {\bf 215}, pp.~203--333, June 1992.

\bibitem{rubakov82}
V.~A. {Rubakov}, M.~V. {Sazhin}, and A.~V. {Veryaskin}, ``{Graviton creation in
  the inflationary universe and the grand unification scale},'' {\em \plb} {\bf
  115}, pp.~189--192, Sept. 1982.

\bibitem{grishchuk75}
L.~P. Grishchuk, ``Amplification of gravitational waves in an istropic
  universe,'' {\em Sov. Phys. JETP} {\bf 40}, pp.~409--415, 1975.

\bibitem{abbott84a}
L.~F. {Abbott} and M.~B. {Wise}, ``{Constraints on generalized inflationary
  cosmologies},'' {\em Nuclear Physics B} {\bf 244}, pp.~541--+, Oct. 1984.

\bibitem{kamionkowski97b}
M.~{Kamionkowski}, A.~{Kosowsky}, and A.~{Stebbins}, ``{A Probe of Primordial
  Gravity Waves and Vorticity},'' {\em \prl} {\bf 78}, pp.~2058--2061, Mar.
  1997.
\newblock astro-ph/9609132.

\bibitem{seljak97}
U.~{Seljak} and M.~{Zaldarriaga}, ``{Signature of Gravity Waves in the
  Polarization of the Microwave Background},'' {\em \prl} {\bf 78},
  pp.~2054--2057, Mar. 1997.
\newblock astro-ph/9609169.

\bibitem{kamionkowski97a}
M.~{Kamionkowski} and A.~{Loeb}, ``Getting around cosmic variance,'' {\em \prd}
  {\bf 56}, p.~4511, Oct. 1997.

\bibitem{zaldarriaga97}
M.~{Zaldarriaga} and U.~{Seljak}, ``All-sky analysis of polarization in the
  microwave background,'' {\em \prd} {\bf 55}, pp.~1830--1840, 1997.

\bibitem{baccigalupi03}
C.~{Baccigalupi}, ``{Cosmic microwave background polarisation: foreground
  contrast and component separation},'' {\em New Astronomy Review} {\bf 47},
  pp.~1127--1134, Dec. 2003.

\bibitem{ball}
D.~{Ball, (NSBF Head of Operations)} 2004.
\newblock Private communication: Average length Antarctica balloon flight of 14
  days.

\bibitem{bacci_private}
C.~{Baccigalupi} 2004.
\newblock Private communication.

\bibitem{hanany03b}
S.~{Hanany} and P.~{Rosenkranz}, ``{Polarization of the atmosphere as a
  foreground for cosmic microwave background polarization experiments},'' {\em
  New Astronomy Review} {\bf 47}, pp.~1159--1165, Dec. 2003.

\bibitem{keating98}
B.~Keating, P.~Timbie, P.~Polnarev, and J.~Steinberger, ``Large angular scale
  polarization of the cosmic microwave background and the feasibility of its
  detection,'' {\em \apj} {\bf 496}, p.~580, 1998.

\bibitem{zaldarriaga98}
M.~{Zaldarriaga} and U.~{Seljak}, ``{Gravitational lensing effect on cosmic
  microwave background polarization},'' {\em \prd} {\bf 58}, p.~23003 (6
  pages), July 1998.

\bibitem{schwan03}
D.~{Schwan}, F.~{Bertoldi}, S.~{Cho}, M.~{Dobbs}, R.~{Guesten}, N.~W.
  {Halverson}, W.~L. {Holzapfel}, E.~{Kreysa}, T.~M. {Lanting}, A.~T. {Lee},
  M.~{Lueker}, J.~{Mehl}, K.~{Menten}, D.~{Muders}, M.~{Myers}, T.~{Plagge},
  A.~{Raccanelli}, P.~{Schilke}, P.~L. {Richards}, H.~{Spieler}, and
  M.~{White}, ``{APEX-SZ a Sunyaev-Zel'dovich galaxy cluster survey},'' {\em
  New Astronomy Review} {\bf 47}, pp.~933--937, Dec. 2003.

\bibitem{spt}
{http://astro.uchicago.edu/spt}.

\bibitem{lee96}
A.~T. {Lee}, P.~L. {Richards}, S.~W. {Nam}, B.~{Cabrera}, and K.~D. {Irwin},
  ``{A Superconducting Bolometer with Strong Electrothermal Feedback},'' {\em
  Applied Physics Letters} {\bf 69}, pp.~1801--1803, Sept. 1996.

\bibitem{lee98}
S.~{Lee}, J.~M. {Gildemeister}, W.~{Holmes}, A.~T. {Lee}, and P.~L. {Richards},
  ``{Voltage-Biased Superconducting Transition-Edge Bolometer with Strong
  Electrothermal Feedback Operated at 370 mK},'' {\em \ao} {\bf 37},
  pp.~3391--3397, June 1998.

\bibitem{gildemeister99}
J.~M. {Gildemeister}, A.~T. {Lee}, and P.~L. {Richards}, ``{A Fully
  Lithographed Voltage-biased Superconducting Spiderweb Bolometer},'' {\em
  Applied Physics Letters} {\bf 74}, pp.~868--870, Feb. 1999.

\bibitem{gildemeister00}
J.~M. {Gildemeister}, A.~T. {Lee}, and P.~L. {Richards}, ``{Monolithic Arrays
  of Absorber-coupled Voltage-biased Superconducting Bolometers },'' {\em
  Applied Physics Letters} {\bf 77}, pp.~4040--4042, Dec. 2000.

\bibitem{gildemeister00b}
J.~M. Gildemeister, {\em Voltage-Biased Superconducting Bolometers for Infrared
  and mm-Waves}.
\newblock PhD thesis, University of California, Berkeley, 2000.

\bibitem{yoon01}
J.~{Yoon}, J.~{Clarke}, J.~M. {Gildemeister}, A.~T. {Lee}, M.~J. {Myers}, P.~L.
  {Richards}, and J.~T. {Skidmore}, ``{Single Superconducting Quantum
  Interference Device Multiplexer for Arrays of Low-Temperature Sensors},''
  {\em Applied Physics Letters} {\bf 78}, pp.~371--373, Jan. 2001.

\bibitem{cunningham02}
M.~F. {Cunningham}, J.~N. {Ullom}, T.~{Miyazaki}, S.~E. {Labov}, J.~{Clarke},
  T.~M. {Lanting}, A.~T. {Lee}, P.~L. {Richards}, J.~{Yoon}, and H.~{Spieler},
  ``High-resolution operation of frequency-multiplexed transition-edge photon
  sensors,'' {\em Applied Physics Letters} {\bf 81}, pp.~159--161, July 2002.

\bibitem{birch84}
J.~R. {Birch} and {Kong Fan Ping}, ``{An interferometer for the determination
  of the temperature variation of the complex refraction spectra of reasonably
  transparent solids at near-millimetre wavelengths},'' {\em Infrared Physics}
  {\bf 24}, pp.~309--314, May 1984.

\bibitem{birch93}
J.~R. {Birch}, ``{Systematic errors in dispersive fourier transform
  spectroscopy in a non-vacuum environment},'' {\em Infrared Physics} {\bf 34},
  pp.~89--93, 1993.

\bibitem{parshin95}
V.~V. {Parshin}, R.~{Heidinger}, B.~A. {Andreev}, A.~V. {Gusev}, and V.~B.
  {Shmagin}, ``{Silicon as an advanced window material for high power
  gyrotrons},'' {\em Int. J. Infrared and Millimeter Waves} {\bf 16},
  pp.~864--877, 1995.

\bibitem{afsar94}
M.~N. {Afsar} and E.~A. {Nichol}, ``{Millimeter wave complex refractive index,
  complex dielectric permittivity and loss tangent of extra high purity and
  compensated silicon},'' {\em Int. J. Infrared and Millimeter Waves} {\bf 15},
  pp.~1181--1188, 1994.

\bibitem{jones88}
T.~J. {Jones} and D.~{Klebe}, ``{A simple infrared polarimeter},'' {\em \pasp}
  {\bf 100}, pp.~1158--1161, Sept. 1988.

\bibitem{platt91}
S.~R. {Platt}, R.~H. {Hildebrand}, R.~J. {Pernic}, J.~A. {Davidson}, and
  G.~{Novak}, ``{100-micron array polarimetry from the Kuiper Airborne
  Observatory - Instrumentation, techniques, and first results},'' {\em \pasp}
  {\bf 103}, pp.~1193--1210, Nov. 1991.

\bibitem{leach91}
R.~W. {Leach}, D.~P. {Clemens}, B.~D. {Kane}, and R.~{Barvainis},
  ``{Polarimetric mapping of Orion using MILLIPOL - Magnetic activity in
  BN/KL},'' {\em \apj} {\bf 370}, pp.~257--262, Mar. 1991.

\bibitem{murray97}
A.~G. {Murray}, R.~{Nartallo}, C.~V. {Haynes}, F.~{Gannaway}, and P.~A.~R.
  {Ade}, ``{An Imaging Polarimeter for SCUBA},'' in {\em ESA SP-401: The Far
  Infrared and Submillimetre Universe},  pp.~405--+, 1997.

\bibitem{church03b}
S.~{Church}, P.~{Ade}, J.~{Bock}, M.~{Bowden}, J.~{Carlstrom}, K.~{Ganga},
  W.~{Gear}, J.~{Hinderks}, W.~{Hu}, B.~{Keating}, J.~{Kovac}, A.~{Lange},
  E.~{Leitch}, O.~{Mallie}, S.~{Melhuish}, A.~{Murphy}, B.~{Rusholme},
  C.~{O'Sullivan}, L.~{Piccirillo}, C.~{Pryke}, A.~{Taylor}, and K.~{Thompson},
  ``{QUEST on DASI: A South Pole CMB polarization experiment},'' {\em New
  Astronomy Review} {\bf 47}, pp.~1083--1089, Dec. 2003.

\bibitem{johnson03b}
B.~R. {Johnson}, M.~E. {Abroe}, P.~{Ade}, J.~{Bock}, J.~{Borrill}, J.~S.
  {Collins}, P.~{Ferreira}, S.~{Hanany}, A.~H. {Jaffe}, T.~{Jones}, A.~T.
  {Lee}, L.~{Levinson}, T.~{Matsumura}, B.~{Rabii}, T.~{Renbarger}, P.~L.
  {Richards}, G.~F. {Smoot}, R.~{Stompor}, H.~T. {Tran}, and C.~D. {Winant},
  ``{MAXIPOL: a balloon-borne experiment for measuring the polarization
  anisotropy of the cosmic microwave background radiation},'' {\em New
  Astronomy Review} {\bf 47}, pp.~1067--1075, Dec. 2003.
\newblock astro-ph/0308259.

\bibitem{polarbear}
{http://bolo.berkeley.edu/polarbear/}.

\bibitem{koester59}
C.~J. {Koester}, ``{Achromatic combinations of half-wave plates},'' {\em \josa}
  {\bf 49}, pp.~405--409, 1959.

\bibitem{pancharatnam55}
S.~{Pancharatnam}, ``{Achromatic combinations of birefringent plates},'' {\em
  Raman Research Inst. Bangalore, Memoir} , 1955.

\bibitem{rivetti87}
A.~{Rivetti}, G.~{Martini}, R.~{Goria}, and S.~{Lorefice}, ``{Turbine flowmeter
  for liquid helium with the rotor magnetically levitated},'' {\em Cryogenics}
  {\bf 27}, pp.~8--11, 1987.

\bibitem{decher93}
R.~{Decher}, P.~N. {Peters}, R.~C. {Sisk}, E.~W. {Urban}, M.~{Vlasse}, and
  D.~K. {Rao}, ``{High temperature superconducting bearing for rocket engine
  turbopumps},'' {\em Applied Superconductivity} {\bf 1}, pp.~1265--1278, 1993.

\bibitem{hull97}
J.~R. {Hull}, ``{Flywheels on a roll},'' {\em Spectrum, IEEE.} {\bf 34},
  pp.~20--25, 1997.

\bibitem{zhang02}
Y.~{Zhang}, Y.~{Postrekhin}, K.~B. {Ma}, and W.~K. {Chu}, ``{Reaction wheel
  with HTS bearings for mini-satellite attitude control},'' {\em Supercond.
  Sci. Technol.} {\bf 15}, pp.~823--825, May 2002.

\bibitem{hanany03}
S.~{Hanany}, T.~{Matsumura}, B.~{Johnson}, T.~{Jones}, J.~R. {Hull}, and K.~B.
  {Ma}, ``{A cosmic microwave background radiation polarimeter using
  superconducting bearings},'' {\em IEEE Transactions on Applied
  Superconductivity} {\bf 13}, pp.~2128--2133, 2003.

\bibitem{hu03}
W.~{Hu}, M.~M. {Hedman}, and M.~{Zaldarriaga}, ``{Benchmark parameters for CMB
  polarization experiments},'' {\em \prd} {\bf 67}, pp.~043004--+, Feb. 2003.
\newblock astro-ph/0210096.

\end{thebibliography}
\bibliographystyle{spiebib}   

\end{document}